\documentclass[aps,reprint, prl, superscriptaddress, showpacs,nobibnotes]{revtex4-1}
\usepackage{graphicx}
\usepackage{xcolor}
\usepackage{soul}
\usepackage{comment}
\usepackage{hyperref}
\usepackage{bm}
\usepackage{braket}
\usepackage{lineno}

\begin{document}

\title[Signatures of polarized chiral spin disproportionation in rare earth nickelates]{Signatures of polarized chiral spin disproportionation in rare earth nickelates}

\author{Jiarui Li}%
\affiliation{Department of Physics, Massachusetts Institute of Technology, Cambridge, Massachusetts 02139, USA}

\author{Robert J. Green}%
\affiliation{Department of Physics and Engineering Physics, University of Saskatchewan, Saskatoon, Saskatchewan S7N 5E2, Canada}
\affiliation{Stewart Blusson Quantum Matter Institute, University of British Columbia, Vancouver, British Columbia V6T 1Z4, Canada}

\author{Claribel Domínguez}%
\affiliation{Department of Quantum Matter Physics, University of Geneva, 24 Quai Ernest-Ansermet, 1121 Geneva 4, Switzerland}

\author{Abraham Levitan}%
\affiliation{Department of Physics, Massachusetts Institute of Technology, Cambridge, Massachusetts 02139, USA}

\author{Yi  Tseng}%
\affiliation{Department of Physics, Massachusetts Institute of Technology, Cambridge, Massachusetts 02139, USA}

\author{Sara Catalano}%
\affiliation{Department of Quantum Matter Physics, University of Geneva, 24 Quai Ernest-Ansermet, 1121 Geneva 4, Switzerland}

\author{Jennifer Fowlie}%
\affiliation{Department of Quantum Matter Physics, University of Geneva, 24 Quai Ernest-Ansermet, 1121 Geneva 4, Switzerland}

\author{Ronny Sutarto}%
\affiliation{Canadian Light Source, Saskatoon, SK S7N 2V3, Canada}

\author{Fanny Rodolakis}%
\affiliation{Advanced Photon Source, Argonne National Laboratory, Lemont, IL 60439, USA}

\author{Lucas Korol}%
\affiliation{Department of Physics and Engineering Physics, University of Saskatchewan, Saskatoon, Saskatchewan S7N 5E2, Canada}

\author{Jessica L. McChesney}%
\affiliation{Advanced Photon Source, Argonne National Laboratory, Lemont, IL 60439, USA}

\author{John W. Freeland}%
\affiliation{Advanced Photon Source, Argonne National Laboratory, Lemont, IL 60439, USA}

\author{Dirk Van der Marel}%
\affiliation{Department of Quantum Matter Physics, University of Geneva, 24 Quai Ernest-Ansermet, 1121 Geneva 4, Switzerland}

\author{Marta Gibert}%
\affiliation{Institute of Solid State Physics, Vienna University of Technology, Vienna, Austria}

\author{Riccardo Comin}%
\affiliation{Department of Physics, Massachusetts Institute of Technology, Cambridge, Massachusetts 02139, USA}

\date{\today  \\~jiaruili@mit.edu, rcomin@mit.edu}
\maketitle


\textbf{
In Rare earth nickelates (RENiO$_3$), electron-lattice coupling drives a concurrent metal-to-insulator and bond disproportionation phase transition whose microscopic origin has long been the subject of active debate. 
Of several proposed mechanisms, 
here we test the hypothesis that pairs of self-doped ligand holes spatially condense to provide local spin moments that are antiferromagnetically coupled to  Ni spins. These singlet-like states provide a basis for long-range bond and spiral spin order. Using magnetic resonant X-ray scattering on NdNiO$_3$ thin films, we observe the chiral nature of the spin-disproportionated state, with spin spirals propagating along the crystallographic (101)$_\mathrm{ortho}$ direction. These spin spirals are found to preferentially couple to X-ray helicity, establishing the presence of a hitherto-unobserved macroscopic chirality. The presence of this chiral magnetic configuration suggests a potential multiferroic coupling between the noncollinear magnetic arrangement and improper ferroelectric behavior as observed in prior studies on NdNiO$_3$ (101)$_\mathrm{ortho}$ films 
and RENiO$_3$ single crystals. 
Experimentally-constrained theoretical double-cluster calculations confirm the presence of an energetically stable spin-disproportionated state with Zhang-Rice singlet-like combinations of Ni and ligand moments.
}

\begin{figure}
	\centering
	\includegraphics[width = 1.0 \linewidth]{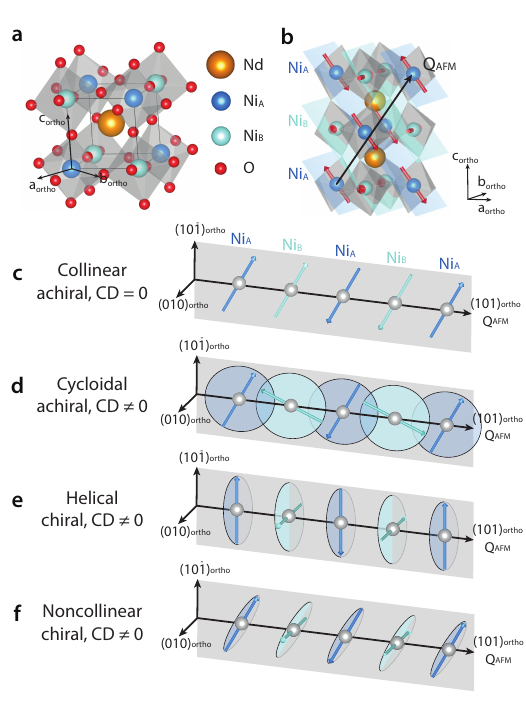}
	\caption{\label{fig_structure}
	\textbf{Crystal structure and possible magnetic states for NdNiO$_3$.} \textbf{a}, Crystal structure of paramagnetic NdNiO$_3$ (NNO). \textbf{b}, Magnetic structure for NNO (101)$_\mathrm{ortho}$, showing   the AFM ordered Ni sublattice and period-four magnetic supercell with ordering vector $\mathbf{Q}_\mathrm{AFM} = $(1/2, 0, 1/2)$_\mathrm{ortho}$, leading to two distinct Ni planes. \textbf{c-f}, Possible magnetic structures that are allowed by magnetic space group (MSG) symmetry. Among them, the collinear (\textbf{c}) and cycloidal (\textbf{d}) magnetic structure are achiral while the helical (\textbf{e}) and general noncollinear (\textbf{f}) magnetic structure are chiral. The X-ray scattering scattering circular dichroism (CD) is expected for all magnetic structures except collinear AFM.
    }
\end{figure}

Systems with strong electron-electron correlations often display ground states with various kinds of symmetries spontaneously broken. In the presence of strong coupling between charge, spin, and lattice, a common instability involves breaking of the translational symmetry of the host crystals with accompanying spatial modulation of the charge/spin density \cite{gruner2000}.
RENiO$_3$ nickelates are host to one such phase of matter with bond/charge disproportionation and a complex noncollinear antiferromagnetic (AFM) magnetic structure \cite{Catalan2008, Middey2016, Catalano2018, Song2022, Torrance1992, Alonso1999, Staub2002, Medarde2009, Peil2019} [Fig. \ref{fig_structure}(a)]. 
The nominal $3d^7$ electron filling at the Ni site is unstable to the formation of holes at the ligand site (oxygen) \cite{Park2012, Mizokawa1999, Lee2011, Johnston2014, Bisogni2016, Varignon2017}. The presence of these ground state oxygen holes has been proposed as the mechanism breaking the symmetry between equivalent Ni atoms, leading to electronically distinct $3d^8$ and $3d^8 \underline{L}^2$ configurations ($\underline{L}$: ligand hole) and setting the stage for a disproportionated state at inequivalent Ni$_{A,B}$ sites. In the extreme case of this scenario, pairs of ligand (oxygen) holes with total spin $S_{\mathrm{lig}} = 1$ antiferromagnetically bind to Ni $S_{\mathrm{Ni}} = 1$ spins every other site, producing an alternation of pristine and suppressed local moments in the expanded  (Ni$_{A}$O$_6$) and compressed (Ni$_{B}$O$_6$) octahedra, respectively. The formation of these singlet-like states and resulting spin disproportionation is central to the description of the ground state of RENiO$_3$. However, it has been experimentally elusive. Moreover, RENiO$_3$ have been proposed to be a type-II multiferroic \cite{Giovannetti2009}. Although the existence of inversion symmetry breaking and potential polar order has been reported within the magnetic phase of NdNiO$_3$ (101)$_\mathrm{ortho}$ film \cite{Kim2016} and RENiO$_3$ single crystals \cite{Ardizzone2021}, the connection to the underlying spin texture of these systems has yet to be investigated.

In this work, we explore the microscopic spin physics of RENiO$_3$ through the lens of its AFM ground state, which is known to emerge within the bond/charge disproportionated insulating phase. We use polarized resonant magnetic X-ray scattering and circular dichroism (CD) to resolve a chiral magnetic ground state with a robust net macroscopic chirality. Using site-resolved X-ray spectra, we find experimental evidence for a spin disproportionated state, which is corroborated by a double-cluster exact diagonalization study confirming the presence of local antiferromagnetically coupled spins at the Ni$_B$ site.

The AFM state of RENiO$_3$ is characterized by a period-four magnetic supercell with inequivalent Ni sublattices (Ni$_\mathrm{A,B}$) and a propagation vector $\mathbf{Q}_\mathrm{AFM}$ = $(1/2, 0, 1/2)_\mathrm{ortho}$ (ortho: orthorhombic notation) as shown in Fig. \ref{fig_structure}(b). Neutron powder diffraction experiments provided evidence for magnetic order but could not resolve whether the magnetic moments arrange in a collinear or noncollinear pattern  \cite{Garcia1994, Fernandez-Diaz, Gawryluk2019}. These studies also revealed the existence of two inequivalent nickel sites with different magnetic moment in Ho, Pr and Nd compounds. Later, polarized resonant magnetic X-ray scattering experiments on single-crystalline thin films have uncovered a prevalence of noncollinear magnetic structures in RENiO$_3$ \cite{Scagnoli2006, Scagnoli2008, Frano2013, Hepting2018}. The refined magnetic structures include collinear AFM, magnetic cycloids, and more general noncollinear magnetic spirals. Although numerous magnetic structures have been identified, the microscopic characteristics of the antiferromagnetic (AFM) state in RENiO$_{3}$ continue to be a topic of intense scholarly discussion.

Insights into the magnetic ground state of RENiO$_3$ can be obtained from magnetic space group (MSG) analysis, which provides the basic symmetry constraints for the Ni moment orientations \cite{Sup}. The four types of possible arrangements of the Ni moment (collinear, cycloidal, helical, and general noncollinear AFM) are shown in Fig. \ref{fig_structure}(c-f). Among all magnetic space groups with maximal magnetic symmetry, only $P_{2a}c$ and $P_{2a}2_1$ are compatible with the noncollinear magnetic structures reported previously\cite{Scagnoli2006, Scagnoli2008, Frano2013, Hepting2018}. The MSG $P_{2a}c$ can host achiral noncollinear magnetic cycloids [Fig. \ref{fig_structure}(d)], whereas the MSG $P_{2a}2_1$ can host a chiral helical magnetic ground state or a more general chiral noncollinear structure [Fig. \ref{fig_structure}(e-f)].

\begin{figure*}
	\centering
	\includegraphics[width = 2 \columnwidth]{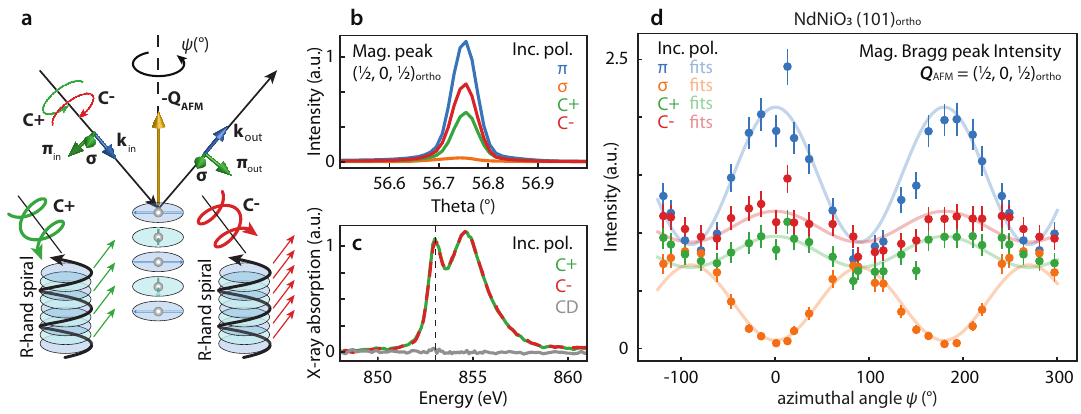}
	\caption{\label{fig_chirality}
	\textbf{X-ray characterization of the magnetic chirality in (101)$_\mathrm{ortho}$-oriented films.} \textbf{a}, Schematic illustration of the experimental geometry. Different polarization states are defined with respect to the incident and scattered photon wavevectors $\mathbf{k}_\mathrm{in,out}$. The axis for the azimuthal rotations of the sample is along $\mathbf{Q}_\mathrm{AFM}$. The bottom left (right) inset illustrates the polarization-dependent scattering of C+ (C-) polarized X-rays by a right-handed magnetic spiral. \textbf{b}, Rocking scans with different incident polarization around the $\mathbf{Q}_\mathrm{AFM} = $ (1/2, 0, 1/2)$_\mathrm{ortho}$ AFM Bragg peak at low temperature (20\,K) and at the Ni $L_3$ resonance ($E$ = 853 eV). A significant circular dichroism is observed. \textbf{c}, X-ray absorption spectra (XAS) across the Ni $L_3$ edge. No circular dichroism (CD, in grey) is found. The vertical dashed-line marks the X-ray energy at which the data in \textbf{b}, \textbf{d} were taken. \textbf{d}, Azimuthal angle $\psi$ dependence of the scattering intensity in different X-ray polarization channels. The error bar on the data points represents the standard deviation of the fit to the rocking curve as in \textbf{b}. The lines represent the results of fit to a noncollinear spiral magnetic structure model described in the text and Supplementary Information \cite{Sup}. All intensities are expressed in arbitrary units (a.u.).
    }
\end{figure*}

The experimental approach hinges on the idea that a noncollinear magnet is  associated with a non-vanishing vector chirality ($\mathbf{\hat{m}_\mathrm{A}} \times \mathbf{\hat{m}_\mathrm{B}}$). Using circular polarized x-rays, we can distinguish between collinear and noncollinear magnetic structures, as well as vector chirality with opposite signs, as illustrated in Fig. \ref{fig_chirality}(a) \cite{Durr1999, Sup}. This distinction is unattainable with linearly polarized X-rays, which have been used to investigate complex magnetic structures in previous studies \cite{Frano2013, Hepting2018, Sears2020}.

The resonant magnetic X-ray scattering scans across the $\mathbf{Q}_\mathrm{AFM} = $ (1/2, 0, 1/2)$_\mathrm{ortho}$ AFM Bragg peak at 20 K in a NNO thin film are shown in Fig. \ref{fig_chirality}(b). The magnetic scattering intensity is maximal for incident $\pi$-polarized X-rays and vanishingly small for $\sigma$-polarized X-rays [the scattering geometry and polarization vectors are defined in Fig. \ref{fig_chirality}(a) and Supplementary Information \cite{Sup}]. At the same time, we observed an unprecedented and substantial circular dichroism of around 25\% ($CD = \frac{2|I_\mathrm{C+} - I_\mathrm{C-}|}{I_\mathrm{C+}+I_\mathrm{C-}}$), which unambiguously points to a  noncollinear magnetic structure. This large CD is exclusive to the X-ray scattering process, and absent in the X-ray absorption spectra across the Ni $L_3$ edge [Fig. \ref{fig_chirality}(c)]. Therefore, we can rule out that the circular dichroic signal observed in the magnetic scattering data arises from a net magnetization or from extrinsic effects in the probing geometry (e.g., uncompensated incident flux for the two photon helicities). These data not only indicate that the microscopic magnetic texture is chiral, but also reflect the presence of a net macroscopic chirality, or an imbalance between domains with opposite vector chirality (if the samples had net zero chirality, the CD contribution from domains of opposite vector chirality would cancel out).

To determine the microscopic magnetic configuration, we collected the magnetic scattering intensity data at various azimuthal angles $\psi$. These angles correspond to rotations of the sample around $\mathbf{Q}_\mathrm{AFM}$, preserving the Bragg's condition while modulating the linear magnetic scattering amplitude $\left( \hat{\mathbf{\epsilon}}^*_\mathrm{out} \times \hat{\mathbf{\epsilon}}_\mathrm{in} \right) \cdot \hat{\mathbf{m}}_i$ (where $\hat{\mathbf{\epsilon}}$ is the X-ray polarization). This approach enables the unambiguous reconstruction of the magnetic structure. The azimuthal angle dependence of the magnetic scattering intensity is shown in Fig. \ref{fig_chirality}(d). The magnetic scattering intensity in all channels exhibits a systematic two-fold modulation upon a full $\psi$ rotation. In the $\pi$-polarization channel, the intensity is maximized at $\psi = 0^\circ$ and $180^\circ$, corresponding to the $(hkh)_\mathrm{ortho}$ scattering plane. Meanwhile, the intensity in the $\sigma$ channel almost diminishes. 
The scattering intensities for the two linear polarization channels reach a maximum (minimum) in the $\sigma$ ($\pi$) channel, respectively. The circular dichroic signal undergoes a similar two-fold modulation with the intensity in the circular left polarization channel (C-) systematically higher than the opposite channel (C+) at all angles. The almost constant circular dichroism implies that the vector chirality ($\mathbf{\hat{m}}_\mathrm{A} \times \mathbf{\hat{m}}_\mathrm{B}$) remains almost unchanged with respect to the incident and scattered X-ray polarizations ($\hat{\mathbf{\epsilon}}^*_\mathrm{in,out}$) upon rotating $\psi$, pointing to a helical magnetic structure.

\begin{figure*}
\includegraphics[width = 150.4 mm] {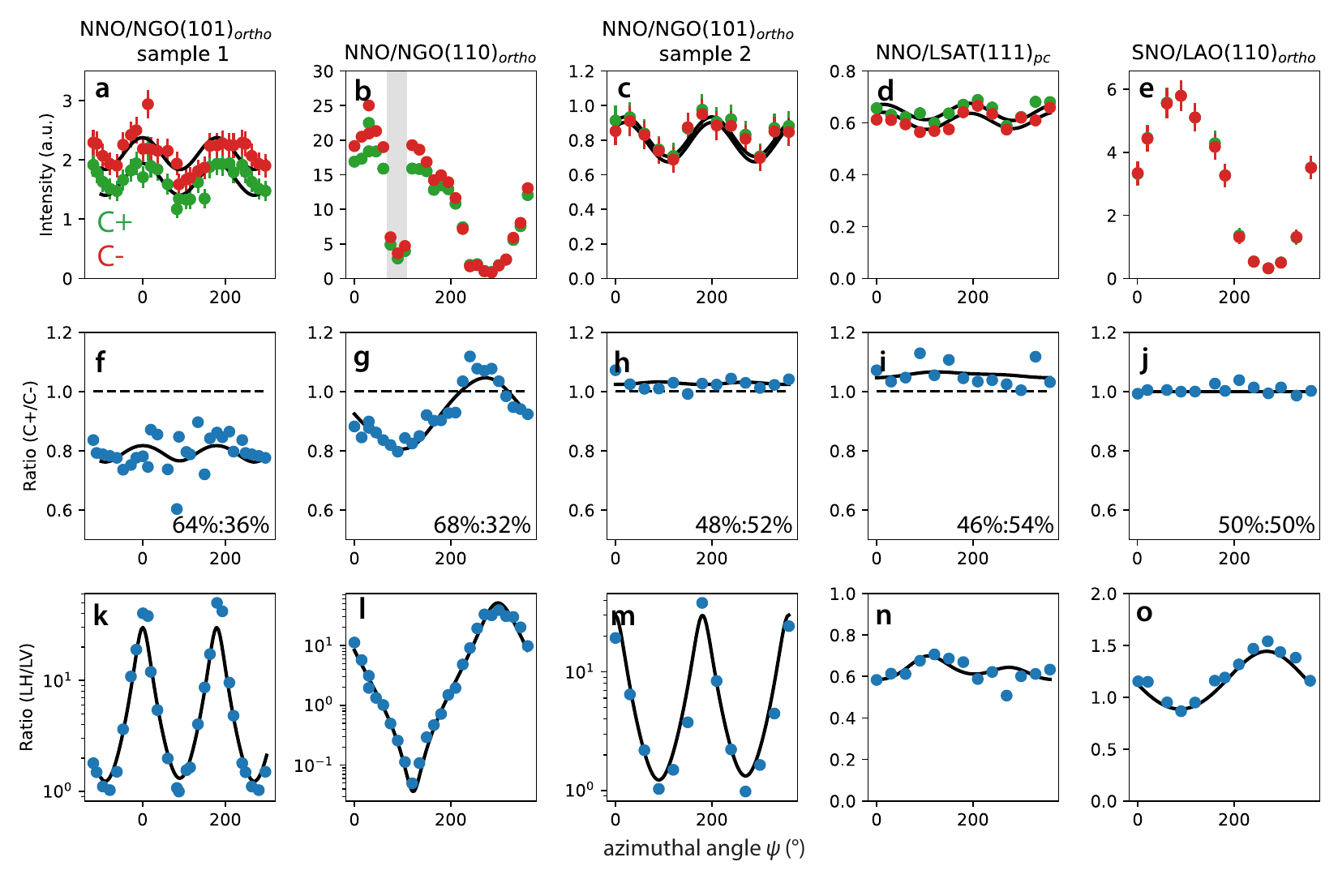}
	\caption{\label{figs_sample_compare} \textbf{Net chirality of magnetic domains in NdNiO$_3$ and SmNiO$_3$.} \textbf{a-e}, Azimuthal angle $\psi$ dependence of the scattering intensity measured by different circular polarizations C+/C- for different samples. At angle around 90$^\circ$ [grey area in (b)], the scattering intensity is strongly suppressed due to extreme grazing angles. \textbf{f-o}, Scattering intensity ratio between the two  circular polarization channels C+/C-  (\textbf{f-j}) and linear polarization channels $\pi / \sigma$ (\textbf{k-o}). All samples are fitted using the same model. The fit results are shown in black in different polarization channels.}
\end{figure*}


\begin{figure}
	\centering
	\includegraphics[width = 1.0 \linewidth]{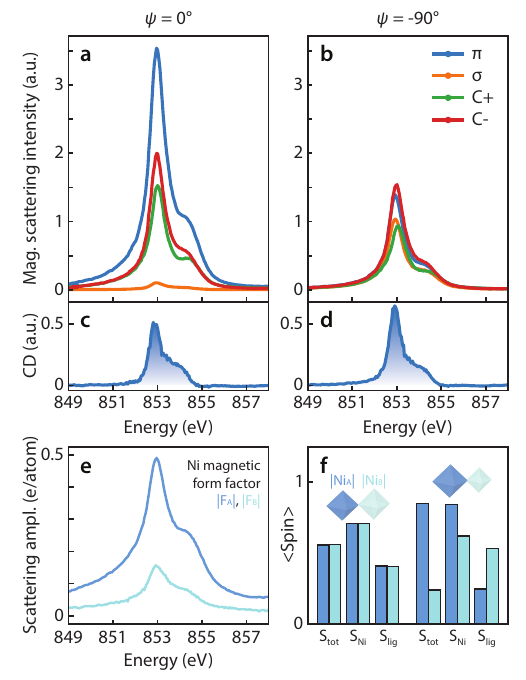}
	\caption{\label{fig_efixQ}
	\textbf{Magnetic scattering form factor for (101)$_\mathrm{ortho}$ NNO film.} \textbf{a},\textbf{b}, Photon energy dependence of the magnetic Bragg peak intensity at $\mathbf{Q}_\mathrm{AFM}$ and across the Ni $L_3$ edge, for azimuthal angles \textbf{a} $\psi = 0^\circ$, and \textbf{b} $\psi = -90^\circ$. The resonant X-ray scattering spectra strongly depend on the X-ray polarization. \textbf{c},\textbf{d}, The circular dichroism (CD) of the resonant X-ray scattering spectra at the azimuthal angles of \textbf{a} and \textbf{b}, respectively. A small difference in spectral lineshape is observed. \textbf{e}, The extracted energy-dependent modulus-squared magnetic scattering form factor $|F_{A,B}(E)|$ for Ni$_{A,B}$ near Ni $L_3$ edge. \textbf{f}, Histogram plot of the expectation values of total ($S_{\mathrm{tot}}$), Ni ($S_{\mathrm{Ni}}$), and ligand ($S_{\mathrm{lig}}$) spins, obtained from a theoretical double-cluster model for the cases of the undistorted (left) and bond-disproportionated (right) structure.
    }
\end{figure}

Coupled with the magnetic space group analysis \cite{Sup}, the signatures of circular dichroism reveal a helical-like structure, which is only compatible with the $P_{2a}2_1$ magnetic space group, where the Ni$_A$ moments are in the $\mathbf{a}_\mathrm{ortho}-\mathbf{c}_\mathrm{ortho}$ plane and Ni$_B$ moments are parallel to $\mathbf{b}_\mathrm{ortho}$. 
To further resolve the microscopic pattern of Ni moments, we simultaneously fit the magnetic peak intensity for all incident X-ray polarizations to a symmetry-constrained model as described in the Supplementary Information \cite{Sup}. This analysis shows that the Ni$_A$ and Ni$_B$ magnetic moments ($\hat{\mathbf{m}}_\mathrm{A}$ and $\hat{\mathbf{m}}_\mathrm{B}$, respectively) are perpendicular to the propagation vector $\mathbf{Q}_\mathrm{AFM}$ and have the following orientations: $\hat{\mathbf{m}}_\mathrm{A}$ is parallel to $(10\overline{1})_\mathrm{ortho}$ and $\hat{\mathbf{m}}_\mathrm{B}$ is almost parallel to  $(010)_\mathrm{ortho}$ . Further, we noticed that a pure right-handed magnetic helix as shown in Fig. \ref{fig_structure}(e) would give rise to a larger circular dichroic signal than was observed. Hence, the model was expanded to incorporate domains with opposite chirality and relative populations $p$ and $1-p$, respectively. The fitting results suggest domain populations of 64$\pm$2$\%$ (36$\pm$2$\%$) for the right-handed (left-handed) helix [fitting results are shown in Fig. \ref{fig_chirality}(d)].

To confirm the robustness of the chiral magnetic ground state, we have examined multiple nickelate samples with different growth orientations where the underlying substrates provide different lattice mismatch, orientation, and strain symmetries \cite{Catalano2015}. Figure \ref{figs_sample_compare} shows the intensity and ratio between the linear and circular polarization channels for four NNO and one SmNiO$_3$ (SNO) films with (101)$_\mathrm{ortho}$ and (110)$_\mathrm{ortho}$ orientations, respectively. All NNO samples exhibit non-vanishing circular dichroism as evidenced by the intensity ratio C+/C- deviating from unity [Fig. \ref{figs_sample_compare}(f-i)], while it is difficult to quantify the presence of a circular dichroic signal above experimental noise level in SNO [Fig. \ref{figs_sample_compare}(e,j)]. These data clearly support a universal noncollinear and chiral magnetic structure in NNO. The fits to the data using the same model give a similar magnetic structure among (101)$_\mathrm{ortho}$ and (110)$_\mathrm{ortho}$ NNOs, respectively, with varying fractional (percentage) population of chiral domains from significant (64:36) to minor (48:52) even in two nominally identical samples [Fig. \ref{figs_sample_compare}(f,h)], suggesting its stochastic nature. However, we note that no change in the domain population ratio was observed upon multiple thermal cycling in single samples and over a period exceeding 20 months. More details on the samples and data are reported in the Supplementary Information \cite{Sup}. 

Without loss of generality, we quantify the relative magnitude of the $\mathbf{\hat{m}_{A,B}}$ moments in NNO/NGO (101)$_\mathrm{ortho}$ by examining \color{black} the energy dependence of magnetic scattering intensity at two high-symmetry azimuthal angles $\psi = 0^\circ$ and $-90^\circ$, where $\mathbf{\hat{m}_{B}}$ and $\mathbf{\hat{m}_{A}}$ are rotated into the scattering plane respectively, suppressing their contribution to the final magnetic scattering intensity at those $\psi$ angles \cite{Sup}. Figure \ref{fig_efixQ}(a,b) shows the energy dependence of the magnetic scattering intensity for different incident polarizations at two selected azimuthal angles. The resonant spectral shape aligns with previous studies with a maximum scattering intensity attained around 853 eV \cite{Scagnoli2006, Bodenthin2011, Frano2013, Hepting2018}.
Intriguingly, the resonant spectra obtained with opposite circularly polarizations display a subtle difference across the Ni absorption edge [Fig. \ref{fig_efixQ}(c,d)], further supporting the hypothesis that the observed chirality arises from noncollinear magnetic helix comprising two inequivalent Ni$_\mathrm{A,B}$ spins.

The resonant spectra at two characteristic azimuthal angles enable us to derive the amplitudes of the site-dependent magnetic scattering factor $|F_\mathrm{A,B}(E)|$, which reflect the multiplet structure associated to Ni spins $\mathbf{\hat{m}_{A,B}}$. Using the formalism detailed in the Supplementary Information, Fig.\,\ref{fig_efixQ}(e) shows the derived magnetic scattering factor amplitude $|F_\mathrm{A,B}(E)|$ from the resonant spectra in Fig. \ref{fig_efixQ}(a,b) \cite{Sup}. We first note that despite the distinct local electronic configurations of the Ni$_\mathrm{A,B}$ sites \cite{Green2016}, the extracted $|F_\mathrm{A,B}(E)|$ have a largely similar resonant spectral shape. 
The scaling factor $|F_\mathrm{A}|$/$|F_\mathrm{B}|$ represents a magnetic moment ratio $|\mathbf{m}_\mathrm{A}|/|\mathbf{m}_\mathrm{B}| = 3.2 \pm 0.4$ suggesting a remarkable difference in the size of the magnetic moments for the two Ni sites. We note that our method of extracting the ratio of magnetic moments $|\mathbf{m}_\mathrm{A}|/|\mathbf{m}_\mathrm{B}|$ descends directly from the experimental data, and therefore it is model-independent. 

To discuss the origin of such a large spin disproportionation across two neighboring Ni spins, we recall that the ground state of RENiO$_3$ is characterized by negative charge transfer energy and a bond-disproportionated state. In this state, the ligand holes are ordered and bound at every other Ni site: ${3d}^{8} \underline{L} + {3d}^{8} \underline{L} \rightarrow {3d}^{8} \underline{L}^{2} + {3d}^{8} $  \cite{Mizokawa1999, Johnston2014}. The strong covalent mixing of Ni $3d$ orbitals with O $2p$ ligand orbitals supports a large antiferromagnetic exchange interaction, promoting the formation of a Zhang-Rice spin singlet-like configuration \cite{Zhang1988} with opposite ligand and Ni spin orientations in compressed ${3d}^{8} \underline{L}^{2}$ octahedral sites \cite{ Park2012}. This interpretation is further supported by our double-cluster calculations. Figure \ref{fig_efixQ}(f) shows the orbital-resolved spin expectation values for compressed vs. expanded NiO$_6$ clusters with and without disproportionation. In the latter case [Fig. \ref{fig_efixQ}(f), left], two octahedra are equivalent, and so are the calculated spins. In the disproportionated case [Fig. \ref{fig_efixQ}(f), right], the expectation values for the Ni ($S_{\mathrm{Ni}}$) and ligand spins ($S_{\mathrm{lig}}$) in the compressed octahedra with ${3d}^{8} \underline{L}^{2}$ electronic configuration converge to around $S\sim0.58$. However, the total spin $S_{\mathrm{tot}}$ is strongly suppressed, confirming the many-body electronic configuration at the Ni$_\mathrm{A}$ octahedron largely projects onto a singlet state.   
On the other hand, the ligand spin in the expanded octahedra (with ${3d}^{8}$ configuration) is suppressed ($\braket{S_{\mathrm{lig}}}\sim0.25$), while the expectation value of the total spin $\braket{ S_{\mathrm{tot}}}$ almost matches the Ni$_A$ spin $\braket{S_{\mathrm{Ni,A}}}$. 

To conclude, we have experimentally observed a macroscopic chirality with large spin disproportionation in RENiO$_3$. The spin disproportionation originates from the antiferromagnetic coupling between Ni and ligand spins, leading to the formation of spatially-modulated Zhang-Rice singlet-like states \cite{Zhang1988}. Our results also have important implications for the understanding of symmetry breaking phenomena in RENiO$_3$. It has been shown by numerical calculations that RENiO$_3$ might be a type-II multiferroic \cite{Giovannetti2009}. 
Only recently, evidence of inversion symmetry breaking due to the possible appearance of an improper ferroelectric ground state upon entering the AFM ordered state was reported \cite{Ardizzone2021}. The presence of ordered microscopic dipoles might induce a noncollinear AFM order with a preferred chirality through spin-dipole (magnetoelectric) coupling. Therefore, the fact that the population of chiral domains is imbalanced suggests a possible macroscopic polarization in ferroelectric domains. 
Despite the apparent imbalance in the population of chiral magnetic domains, further investigation is required to understand the nature of the spin-lattice interaction, determine the magnitude and direction of the ferroelectric polarization, and explore the magnetoelectric coupling.

\section{Methods}

The coherently strained NdNiO$_3$ thin films samples (around 27 nm thick) were grown on orthorhombic NdGaO$_3$ substrates with different orientations [$(101)_\mathrm{ortho}$ and $(110)_\mathrm{ortho}$] by off-axis radiofrequency magnetron sputtering as described elsewhere \cite{Scherwitzl2010, Catalano2015}.  NdNiO$_3$  layer thicknesses, thin film quality and metal-insulator transitions were characterized by X-ray diffraction, atomic force microscopy and transport measurements \cite{Sup}. 

Resonant X-ray scattering measurements were performed at the REIXS (10ID-2) beamline of the Canadian Light Source, and at beamline 29-ID-D of the Advanced Photon Source. The photon energy of the X-ray was tuned at 853 eV with incident polarization selected to be $\sigma, \pi$, and circular +/- (C+/C-) unless otherwise noted. All measurements are performed at a base temperature of 20 K. The horizontally scattered photons were collected by a channeltron electron multiplier after a 0.5 mm-wide slit which gives an angular resolution of 0.09$^\circ$. The measurements were carried out by collecting magnetic scattering scans for different X-ray polarizations and different values of azimuthal angles. The X-ray absorption spectra were measured during the same experiment in total fluorescence/electron yield mode.

\section{Data availability}

Data associated with this paper are available on the Harvard Dataverse at https://doi.org/10.7910/DVN/WGKTQ5

\section{Acknowledgement}

We thank G.A. Sawatzky and J.M. Triscone for fruitful discussion. This material is based upon work supported by the Department of Energy, Office of Science, Office of Basic Energy Sciences, under Award Number DE-SC0019126 (X-ray scattering measurements and form factor analysis). RJG was supported by the Natural Sciences and Engineering Research Council of Canada (NSERC). M.G. acknowledges the Swiss National Science Foundation (SNSF) under Project No. PP00P2$\_$170564. Part of the research described in this paper was performed at the Canadian Light Source, a national research facility of the University of Saskatchewan, which is supported by the Canada Foundation for Innovation (CFI), NSERC, the National Research Council (NRC), the Canadian Institutes of Health Research (CIHR), the Government of Saskatchewan, and the University of Saskatchewan. This research used resources of the Advanced Photon Source, a U.S. Department of Energy (DOE) Office of Science user facility at Argonne National Laboratory and is based on research supported by the U.S. DOE Office of Science-Basic Energy Sciences, under Contract No. DE-AC02-06CH11357. The work of DvdM was supported by the Swiss National Science Foundation through project 200020-179157. The authors acknowledge the MIT SuperCloud and Lincoln Laboratory Supercomputing Center for providing computing resources that have contributed to the research results reported within this paper.

\section{Author contributions}

J.L. and R.C. conceived the project.
The samples were grown by C.D., J.F. and S.C.
J.L., A.L. and Y.T. performed the experiments with assistance of R.S., F.R., J.M. and J.W.F.
J.L. analysed the data with contributions from R.G., D.V.d.M.
J.L., L.K., and R.G carried out the calculations.
J.L. and R.C. wrote the manuscript with collaborative contributions from all authors.

\section{Competing interests}

The authors declare no competing interests.

\bibliography{NNO_chiral}

\end{document}